\documentstyle[prb,multicol,aps,graphicx]{revtex}
\begin{document}
\draft
\title{Nucleus-mediated spin-flip transitions in $GaAs$ quantum dots}
\author{Sigurdur I. Erlingsson$^1$, Yuli V.\ Nazarov$^1$, and Vladimir I.\
        Fal'ko$^2$} 
\address{$^{1}$Delft University of Technology, Department of Applied Physics,
         Lorentzweg 1, 2628 Delft, The Netherlands}
\address{$^{2}$School of Physics and Chemistry, Lancaster University,
  Lancaster, LA1 4YB, United Kingdom}
\date{\today}
\maketitle

\begin{abstract}
Spin-flip rates in $GaAs$ quantum dots can be quite slow, thus opening up
the possibilities to manipulate spin states in the dots. We present here
estimations of inelastic spin-flip rates mediated by hyperfine interaction
with nuclei. Under general assumptions the nucleus mediated rate is
proportional to the phonon relaxation rate for the corresponding
non-spin-flip transitions.  The rate can be accelerated in the vicinity of
a singlet-triplet excited states crossing. The small proportionality
coefficient depends inversely on the number of nuclei in the quantum dot. We
compare our results with known mechanisms of spin-flip in $GaAs$ quantum dot.
\end{abstract}
\begin{multicols}{2}

The electron spin states in bulk semiconductor and heterostructures have
attracted much attention in recent years. Experiments indicate very long
spin decoherence times and small transition rates between states of
different spin \cite{kikkawa98:4313,ohno00:817,fujisawa01:R81304}. These
promising results have motivated proposals for information processing based
on electron spins in quantum dots, which might lead to a realization of a
quantum computer \cite{loss98:120,burkard99:2070}.

A quantum dot is region where electrons are confined. The energy spectrum is
discrete, due to the small size, and can display atomic like properties \cite
{ashoori96:413,tarucha96:3613}. Here we will consider quantum 
dots in $GaAs$-$AlGaAs$ heterostructures. The main reasons for studying them
are that relevant quantum dots are fabricated in such structures and $GaAs$
has a peculiar electron and phonon properties which are of interest.
There are two main types of gate controlled dots in these systems, so-called
vertical and lateral dots\cite{sohn97:105}. They are characterized by
different transverse confinement, which is approximately a triangular well
and a square well for the lateral and vertical dots, respectively.

Manipulation the electron spin in a coherent way requires that it should be
relatively well isolated from the surrounding environment. Coupling a
quantum dot, or any closed quantum system, to its environment can cause
decoherence and dissipation. One measure of the strength of the coupling to
the environment are the transition rates, or inverse lifetimes, of the
quantum dot states. Calculations of transition rates between different spin
states due to phonon-assisted spin-flip process mediated by spin-orbit
coupling, which is one possibility for spin relaxation, have given
surprisingly low rates in quantum dots \cite
{khaetskii00:470,khaetskii00:12639,frenkel91:14228}. For these calculations
it is very important that the electron states are discrete, and the result
differs strongly from that obtained in application to 2D extended electron
states in $GaAs$. The same argument applies to the phonon scattering
mechanisms, since certain phonon processes possible in 2D and 3D electron
systems are not effective in scattering the electron in 0D. An alternative
mechanism of spin relaxation in quantum dots is caused by hyperfine coupling
of nuclear spins to those of electrons. Although the hyperfine interaction
mediated spin relaxation in donors was considered a long time ago \cite
{pines57:489}, no analysis has been made, yet, of the hyperfine interaction
mediated spin flip processes in quantum dots.

The present publication offers an estimation of the scale of hyperfine
interaction induced spin relaxation rates in $GaAs$ quantum dots and its
magnetic field dependence, the main result presented by the expression in
Eq.\ (\ref{eq:rateHyperfine}). Since the parameters of hyperfine interaction
between conduction band electrons and underlying nuclei in $GaAs$ have been
extensively investigated \cite{paget77:5780,meier84:381}, including the
Overhauser effect and spin-relaxation in $GaAs/AlGaAs$ heterostructures \cite
{dobers88:1650,berg90:2563,vagner82:635,barret95:5112,tycko95:1460} , we are
able now to predict the typical time scale for this process in particular
quantum dot geometries. The rate that we find depends inversely on the
number of nuclei in the dot (which can be manipulated by changing the gate
voltage) and is proportional to the inverse squared exchange splitting in
the dot (which can be varied by application of an external magnetic field
with orientation within the 2D plane). The following text is organized in
two sections: section \ref{sectionII}, where the transition rates in systems
with discrete spectra are analyzed and section \ref{sectionIII} where the
obtained result is compared to transition rates provided by the spin-orbit
coupling mechanism.

\section{Model and assumptions}
\label{sectionII} 

\begin{figure}
\includegraphics*[angle=0,width=8cm]{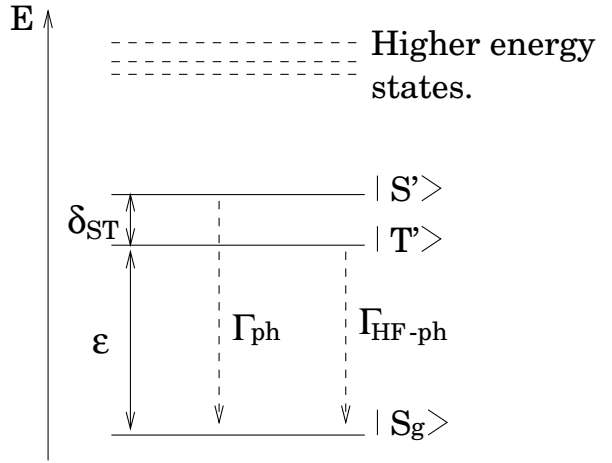}
\caption{The lowest lying states of the quantum dot. The energy separation
of the two singlet states is denoted by $\varepsilon$ and the
exchange splitting by $\delta_{ST}$. The two rates indicated are the
phonon rate $\Gamma_{\scriptsize \mbox{ph}}$ and the combined hyperfine and
phonon rate $\Gamma_{\scriptsize \mbox{HF-ph}}$}
\label{fig:fig1}
\end{figure}

The ground state of a quantum dot is a many electron
singlet $|S_g \rangle$, for sufficiently low magnetic fields. This can
change at higher magnetic fields. We assume that the system is in a regime
of magnetic field so that the lowest lying states are ordered as shown in
Fig.~\ref{fig:fig1}. The relevant energy scales used in the following
analysis are given by the energy difference between the triplet state (we
assume small Zeeman splitting) and the ground state $\varepsilon
=E_{T^{\prime }}-E_{g}$, and exchange splitting $\delta_{ST}=E_{S^{\prime
}}-E_{T^{\prime }}$ between the first excited singlet and the triplet. It is
possible to inject an electron into an excited state of the dot. If this
excited state is a triplet, the system may get stuck there since a spin-flip
is required to cause transitions to the ground state.

The $\Gamma$-point of the conduction band in $GaAs$ is mainly composed of $s$%
-orbitals, so that the hyperfine interaction can be described by the contact
interaction Hamiltonian \cite{brown98:49} 
\begin{equation}
H_{{\scriptsize \mbox{HF}}}=A\sum_{i,k}\bbox{S}_{i}\cdot \bbox{I}%
_{k}\,\delta (\bbox{r}_{i}-\bbox{R}_{k})  \label{eq:hyperfineHamiltonian}
\end{equation}
where $\bbox{S}_{i}$ ($\bbox{I}_{k}$) and $\bbox{r}_{i}$ ($\bbox{R}_{k}$)
denote the spin and position the $i$-th electron ($k$-th nuclei). This
coupling flips the electron spin and simultaneously lowers/raises the $z$%
-component of a nuclear spin, which mixes spin states and provides
possibility for the relaxation.

But the hyperfine interaction alone does not guarantee that transitions
between the above-described states occurs, since the nuclear spin flip
cannot relax the excessive initial state energy. (The energy associated with
a nuclear spin is the nuclear Zeeman, $\hbar \omega _{n}$, energy which is
three orders of magnitude smaller than the electron Zeeman energy and the
energies related to the orbital degree of freedom.) For free electrons, the
change in energy accompanying a spin flip caused by the hyperfine scattering
is compensated by an appropriate change in its kinetic energy. In the case
of a quantum dot, or any system with discrete a energy spectrum, this
mechanism is not available and no hyperfine induced transitions will occur
because the energy released by the quantum can not be absorbed. Therefore,
the spin relaxation process in a dot also requires taking into account the
electron coupling to the lattice vibrations. The excess energy from the
quantum dot can be emitted in the form of a phonon. Since the `bare'
electron-phonon interaction, $H_{\scriptsize \mbox{ph}}$ does not contain any
spin operators and thus does not couple directly different spin states, one
has to employ second order perturbation theory which results in transitions
via virtual states. The amplitude of such a transition between the triplet
state $|T^{\prime }\rangle $ and the ground state $|S_{g}\rangle $ is  
\begin{eqnarray}
\langle T^{\prime }|S_{g}\rangle &=&\hspace{0.2cm}\sum_{t}\frac{\langle
T^{\prime }|H_{{\scriptsize \mbox{ph}}}|t\rangle \langle t|H_{{\scriptsize %
\mbox{HF}}}|S_{g}\rangle }{E_{T^{\prime }}-(E_{t}+\hbar \omega _{\bbox{q}})}
\nonumber \\
&&+\sum_{s}\frac{\langle T^{\prime }|H_{{\scriptsize \mbox{HF}}}|s\rangle
\langle s|H_{{\scriptsize \mbox{ph}}}|S_{g}\rangle }{E_{T^{\prime
}}-(E_{s}+\hbar \omega _{n})}
\end{eqnarray}
where $\hbar \omega _{\bbox{q}}$ is the energy of the emitted phonon and $%
\hbar \omega _{n}\approx 0$ is the energy changed by raising/lowering a
nuclear spin.

It is natural to assume that the exchange splitting is smaller than the
single-particle level splitting, so that the dominating contribution to the
amplitude $\langle T^{\prime }|S_{g}\rangle $ comes from the term describing
the virtual state $|s\rangle =|S^{\prime }\rangle $ , due to a small
denominator. All other term can be ignored and the approximate amplitude
takes the form 
\begin{equation}
\langle T^{\prime }|S_{g}\rangle \approx \frac{\langle T^{\prime }|H_{%
{\scriptsize \mbox{HF}}}|S^{\prime }\rangle \langle S^{\prime }|H_{%
{\scriptsize \mbox{ph}}}|S_{g}\rangle }{E_{T^{\prime }}-E_{S^{\prime }}}.
\label{eq:amplitude}
\end{equation}
The justification for this assumption is that we aim at obtaining estimates
of the rates, and including higher states would not affect the order of
magnitude, even if the exchange splitting is substantial. Note that the
phonon and nuclear state are not explicitly written in Eq.\ (\ref
{eq:amplitude}).

The transition rate from $|T^{\prime}\rangle$ to $|S_g \rangle$ is given by
Fermi's golden rule 
\begin{eqnarray}
\tilde{\Gamma}_{{\scriptsize \mbox{HF-ph}}}&=& \frac{2\pi}{\hbar}
\sum_{N^{\prime}_{\bbox{q}},\bbox{\mu}^{\prime}} |\langle T^{\prime}|S_g
\rangle |^2 \delta(E_{{\scriptsize \mbox{i}}}-E_{{\scriptsize \mbox{f}}})
\label{eq:goldenRule}
\end{eqnarray}
where $N^{\prime}_{\bbox{q}}$ and $\bbox{\mu}^{\prime}$ are the final phonon
and nuclear states respectively and $E_{\scriptsize \mbox{i}}$ and
$E_{\scriptsize \mbox{f}}$ stand for the initial and final energies. Inserting
Eq.\ (\ref{eq:amplitude}) into (\ref{eq:goldenRule}) and averaging over
initial nuclear states with probability $P(\bbox{\mu})$, we obtain an
approximate equation for the nucleus mediated transition rate 
\begin{eqnarray}
\Gamma_{{\scriptsize \mbox{HF-ph}}}&=& \frac{2\pi}{\hbar}\sum_{N^{\prime}_{%
\bbox{q}}} |\langle S^{\prime};N^{\prime}_{\bbox{q}}|H_{{\scriptsize %
\mbox{ph}}} |S_g;N_{\bbox{q}} \rangle |^2 \delta(E_{{\scriptsize \mbox{i}}%
}-E_{{\scriptsize \mbox{f}}})   \nonumber \\
& & \times \sum_{\bbox{\mu}^{\prime},\bbox{\mu}} \frac{P(\bbox{\mu})|\langle
T^{\prime};\bbox{\mu}^{\prime}| H_{{\scriptsize \mbox{HF}}}|S^{\prime};%
\bbox{\mu} \rangle |^2}{(E_{T^{\prime}}-E_{S^{\prime}})^2} \\
&=& \Gamma_{{\scriptsize \mbox{ph}}}(\varepsilon) \sum_{\bbox{\mu}^{\prime},%
\bbox{\mu}} \frac{P(\bbox{\mu})|\langle T^{\prime};\bbox{\mu}^{\prime}| H_{%
{\scriptsize \mbox{HF}}}|S^{\prime};\bbox{\mu} \rangle |^2}{%
(E_{T^{\prime}}-E_{S^{\prime}})^2}  \label{eq:rateDefinition}
\end{eqnarray}
where $\Gamma_{\scriptsize \mbox{ph}}$ is the non-spin-flip phonon rate as a
function of the relaxed energy $\varepsilon=E_{T^{\prime}}-E_{S_g}$.

We will approximate the many-body orbital wavefunctions by symmetric, $|\Psi
_{S}\rangle $, and antisymmetric, $|\Psi _{T}\rangle $, \ Slater
determinants corresponding to the singlet and triplet states respectively.
It is not obvious {\it a priori} why this approximation is applicable, since
the Coulomb interaction in few electron quantum dots can be quite strong 
\cite{daniela01:01}. The exact energy levels are very different from those
obtained by simply adding the single particle energies. However the
wavefunction will not drastically change and especially the matrix elements
calculated using Slater determinants are comparable to the ones obtained by
using exact ones. The singlet and triplet wavefunctions can be decomposed
into orbital and spin parts: $|T^{\prime }\rangle =|\Psi _{T}\rangle
|T\rangle $ and $|S\rangle =|\Psi _{S}\rangle |S\rangle $, where 
\begin{equation}
|T\rangle =\frac{\nu _{x}+i\nu _{y}}{\sqrt{2}}|1,+1\rangle +\frac{\nu
_{x}-i\nu _{y}}{\sqrt{2}}|1,-1\rangle +\nu _{z}|1,0\rangle
\label{eq:expansion}
\end{equation}
and the hyperfine contribution to the rate becomes 
\begin{eqnarray}
\sigma _{{\scriptsize \mbox{HF}}} &=&\sum_{\bbox{\mu'}\bbox{\mu}}P(\bbox{\mu}%
)|\langle T^{\prime };\bbox{\mu}|H_{n}|S^{\prime };\bbox{\mu'}\rangle |^{2} 
\nonumber \\
&=&\frac{A^{2}}{4}G_{{\scriptsize \mbox{corr}}}\sum_{k}\left( |\Psi _{1}(%
\bbox{R}_{k})|^{2}-|\Psi _{2}(\bbox{R}_{k})|^{2}\right) ^{2}
\label{eq:prefactor1}
\end{eqnarray}
where $\Psi _{1;2}$ are the wavefunctions of the lowest energy states and $%
G_{\scriptsize \mbox{corr}}$ contains the nuclear correlation functions 
\begin{equation}
G_{{\scriptsize \mbox{corr}}}=\sum_{\eta,\gamma=x,y,z }\nu _{\eta }\nu
_{\gamma }^{\ast }\left( \bar{G}_{\eta \gamma }+\frac{1}{2}i\epsilon _{\eta
\gamma \kappa }\langle I_{\kappa }\rangle \right) .  \label{eq:corr}
\end{equation}
where $\bar{G}_{\eta \gamma }$ is the symmetric part of the nuclear
correlation tensor, $\epsilon _{\eta \gamma \kappa}$ is the totally
anti-symmetric tensor and the $\nu_\eta$'s are the coefficients in the
triplet state expansion in Eq.\ (\ref{eq:expansion}). We assume that nuclei
are identical and non-interacting, which gives $G_{\scriptsize
  \mbox{corr}}=1.25$ for an isotropic nuclear system. 

Let us now introduce the length scales $\ell $ and $z_{0}$ which are the
spatial extent of the electron wavefunction in the lateral direction and the
dot thickness respectively. Let $C_{n}$ denote concentration of nuclei with
non-zero spin. The effective number of nuclei contained within the quantum
dot is 
\begin{equation}
N_{{\scriptsize \mbox{eff}}}=C_{n}\ell ^{2}z_{0}.  \label{eq:Neff}
\end{equation}
In $GaAs$ $N_{\scriptsize \mbox{eff}}\gg 1$ and the sum over the nuclei in
Eq.\ (\ref{eq:prefactor1}) can be replaced by $C_{n}\int d^{3}\bbox{R}_{k}$
and we define the dimensionless quantity 
\begin{equation}
\gamma _{{\scriptsize \mbox{int}}}=\ell ^{2}z_{0}\int d^{3}\bbox{R}%
_{k}\left( |\Psi _{1}(\bbox{R}_{k})|^{2}-|\Psi _{2}(\bbox{R}%
_{k})|^{2}\right) ^{2}  \label{eq:gammaInt}
\end{equation}
To relate the hyperfine constant $A$ (which has dimension Energy$\times $%
Volume) to more convenient parameter we note that the splitting of spin up
and spin down states at maximum nuclear polarization is $E_{n}=AC_{n}I$,
where $I=3/2$ is the nuclear spin. Thus, the hyperfine mediated transition
rate is 
\begin{equation}
\Gamma _{{\scriptsize \mbox{HF-ph}}}=\Gamma _{{\scriptsize \mbox{ph}}%
}(\varepsilon )\left( \frac{E_{n}}{\delta _{ST}}\right) ^{2}\frac{G_{%
{\scriptsize \mbox{corr}}}\gamma _{{\scriptsize \mbox{int}}}}{(2I)^{2}N_{%
{\scriptsize \mbox{eff}}}}
\label{eq:rateHyperfine}
\end{equation}
Note that the rate is inversely proportional to the number of nuclei
$N_{\scriptsize \mbox{eff}}$ in the quantum dot and depends on the inverse
square of $\delta_{ST}$, which are both possible to vary in experiments\cite
{vanderwiel98:173,tarucha98:112}.  In particular, the
singlet-triplet splitting of excited states of a dot can be brought down to
zero value using magnetic field parallel to the 2D plane of the
heterostructure, which would accelerate the relaxation process.
The nuclear correlation functions in $G_{\scriptsize \mbox{corr}}$ may also be 
manipulated by optical orientation of the nuclear system\cite
{paget77:5780,meier84:381}.

\section{Comparison and estimates}

\label{sectionIII} We now consider Eq.\ (\ref{eq:rateHyperfine}) for a
specific quantum dot structure. It is assumed that the lateral confinement
is parabolic and that the total potential can be split into a lateral and
transverse part. For vertical dots the approximate transverse wavefunction
is 
\begin{equation}
\chi^{{\scriptsize \mbox{ver}}}(z)= \left ( \frac{2}{z_0} \right )^{1/2}
\sin \left ( \frac{\pi z}{z_0} \right )  \label{eq:wfVertical}
\end{equation}
where $z_0$ is the thickness of the quantum well, i.e.\ the dot thickness.
The wavefunctions in the lateral direction are the Darwin-Fock solutions $%
\phi_{n,l}(x,y)$ with radial quantum number $n$ and angular momentum $l$.
The single particle states corresponding to $(n,l)=(0,0)$ and $(0,\pm 1)$
are used to construct the Slater determinant for a two-electron quantum dot.
In the case of these states the factor $\gamma_{\scriptsize \mbox{int}}$ in
Eq.\ (\ref{eq:gammaInt}) is then $\gamma_{\scriptsize \mbox{int}}=0.12$ for a
vertical dot and $\gamma_{\scriptsize \mbox{int}}=0.045$ for a lateral one.

One property of the Darwin-Fock solution is the relation $\ell^{-2}=\hbar
\omega \, m^*/\hbar^2$ where $\omega=\sqrt{\Omega_0^2+\omega_c^2/4}$ is the
effective confining frequency and $\omega_c=eB/m^*$ the cyclotron frequency.
Inserting this and Eq.\ (\ref{eq:Neff}) into (\ref{eq:rateHyperfine}) the
rate for parabolic quantum dots becomes 
\begin{equation}
\Gamma_{{\scriptsize \mbox{HF-ph}}} =\Gamma_{{\scriptsize \mbox{ph}}%
}(\varepsilon) \hbar\omega \left (\frac{E_n}{\delta_{ST}} \right )^2 \frac{%
G_{{\scriptsize \mbox{corr}}} \gamma_{{\scriptsize \mbox{int}}}}{(2 I)^2}
\left ( \frac{m^*}{\hbar^2 C_n z_0} \right ).  \label{eq:rateHyperfineDF}
\end{equation}

Spin relaxation due to spin orbit related mechanisms in $GaAs$ quantum dots
were investigated by Khaetskii and Nazarov in Refs.\ %
\onlinecite{khaetskii00:12639} and \onlinecite{khaetskii00:470}. We will
summarize their results here for comparison with our hyperfine-phonon
mechanism. In Ref \onlinecite{khaetskii00:12639}, it has been found that the
dominating scattering mechanism is due to the absence of inversion symmetry.
There are three parameters related to this mechanism, 
\begin{mathletters}
\label{eq:rate}
\begin{eqnarray}
\Gamma _{1} &=&\Gamma _{{\scriptsize \mbox{ph}}}(\varepsilon )\frac{8}{3}%
\left( \frac{m^{\ast }\!\beta ^{2}}{\hbar \Omega _{0}}\right) ^{3}
\label{eq:ratea} \\
\Gamma _{5} &=&\Gamma _{{\scriptsize \mbox{ph}}}(\varepsilon )\,6(m^{\ast
}\!\beta ^{2})(g^{\ast }\mu _{{\scriptsize \mbox{B}}}B)^{2}\frac{\hbar
\omega }{(\hbar \Omega _{0})^{4}}  \label{eq:rateb} \\
\Gamma _{2} &=&\Gamma _{{\scriptsize \mbox{ph}}}(\varepsilon )\frac{7}{24}%
\frac{(m^{\ast }\beta ^{2})(\hbar \omega )}{E_{z}^{2}}.  \label{eq:ratec}
\end{eqnarray}
\end{mathletters}
Here, $m^{\ast }\!\beta ^{2}$ is determined by the transverse confinement
and band structure parameters and $E_{z}=\langle p_{z}^{2}\rangle /(2m^{\ast
})$. For vertical dots of thickness $z_{0}=15$\thinspace nm, then $m^{\ast
}\!\beta ^{2}\approx 4\times 10^{-3}$\thinspace meV, but one should be
cautious when considering a different thicknesses, since $m^{\ast }\beta
^{2}\propto z_{0}^{-4}$, so the rates are sensitive to variations in $z_{0}$%
. The transition rates in Eqs.\ (\ref{eq:rateHyperfine}) and (\ref{eq:ratea}%
)-(\ref{eq:ratec}) are all proportional to the phonon rate
$\Gamma_{\scriptsize \mbox{ph}}$ evaluated for the same energy difference
$\varepsilon $. It is thus sufficient to compare the only the proportionality
coefficients. 

Let us now consider for which confining energies the different rates are
comparable. At zero magnetic field $\Gamma_1=\Gamma_5$ at $\hbar \Omega_0
\approx 0.8$\,meV. The estimated confining energies of vertical quantum dots
used in experiment are in the range $(2-5.5)$\,meV\cite
{tarucha96:3613,kouwenhoven97:1788,tarucha98:112}. For those dots $\Gamma_2
\gg \Gamma_1$ due to the very different dependence on the confinement, $%
\Gamma_1 \propto (\hbar\Omega_0)^{-3} $ and $\Gamma_2 \propto \hbar\Omega_0$%
. Doing the same for $\Gamma_1$ and $\Gamma_{\scriptsize \mbox{HF-ph}}$ we
obtain that those rates are equal at $\hbar \Omega_0 \approx 4.4$\,meV. The
numerical values used for the hyperfine rate in Eq.\ (\ref{eq:rateHyperfine})
are the following: $E_n=0.13$\,meV\cite{dobers88:1650},
$\delta_{ST}=2.3$\,meV\cite{bolton96:4780} and a $z_0=15$\,nm. Thus, for
$B=0$\,T and the previously cited experimental values for the confining energy
the dominant transition rate is Eq.\ (\ref{eq:ratec}).

For clarity we will give the values of the rates. The non-spin-flip rate $%
\Gamma _{\scriptsize \mbox{ph}}(\varepsilon )$ is given in KN and using the
values $\hbar \Omega _{0}=5.5$\thinspace meV and $B=0$\thinspace T we get 
\begin{mathletters}
\label{eq:estimate}
\begin{eqnarray}
\Gamma _{{\scriptsize \mbox{ph}}} &\approx &3.6\times 10^{7}\,\mbox{s}^{-1}
\label{eq:estimatea} \\
\Gamma _{{\scriptsize \mbox{HF-ph}}} &\approx &2\times 10^{-2}\,\mbox{s}^{-1}
\label{eq:estimateb} \\
\Gamma _{2} &\approx &1\times 10^{2}\,\mbox{s}^{-1}.  \label{eq:estimatec}
\end{eqnarray}
\end{mathletters}
The values of level separations are $\varepsilon =2.7$\thinspace meV and $%
\delta _{ST}=2.84$\thinspace meV are taken from Ref.\ %
\onlinecite{tarucha98:112}.

An application of a magnetic field to the dot may result in two effects. The
rate $\Gamma _{5}$ becomes larger than $\Gamma _{2}$ for magnetic fields
around $B=1$\thinspace T ($\hbar \Omega _{0}=3$\thinspace meV) to $B\approx
5.4$\thinspace T ($\hbar \Omega _{0}=5.5$\thinspace meV). More importantly,
the exchange splitting $\delta _{ST}$ can vanish in some cases and the
hyperfine rate in Eq.\ (\ref{eq:rateHyperfine}) will dominate. It is also
worth noting that in this limit the approximation used in obtaining Eq.\ (%
\ref{eq:amplitude}) becomes very good. Since the rates considered here are
all linear in the phonon rate the divergence of $\Gamma_{\scriptsize
  \mbox{ph}}$ at the singlet-triplet transition does not affect the ratio of
the rates. The above estimates were focused on vertical dots. To obtain the
corresponding results for lateral dots the value of $\gamma_{\scriptsize
  \mbox{int}}$ in Eq.\ (\ref{eq:rateHyperfineDF}) should be used. 

In summary, we have calculated the nucleus mediated spin-flip transition
rate in $GaAs$ quantum dots. The comparison of our results to those
previously obtained for spin-orbit scattering mechanism indicates that the
rates we obtained here are relatively low, due to the discrete spectrum, so
that we believe that hyperfine interaction would not cause problems for
spin-coherent manipulation with $GaAs$ quantum dots. Nevertheless, the
hyperfine rate, which was found to be lower than the spin-orbit rates at
small magnetic field, may diverge and become dominant at certain values
of magnetic field corresponding to the resonance between triplet and
singlet excited states in the dot.

This work is a part of the research program of the ``Stichting vor
Fundementeel Onderzoek der Materie (FOM)'', EPSRC and INTAS. One of authors
(VF) acknowledges support from NATO CLG and thanks R. Haug and I. Aleiner
for discussions.

\end{multicols}
\end{document}